\documentclass[12pt,preprint]{aastex}

\accepted{}

\shorttitle{HIERARCHICAL OBJECT FORMATION IN PECULIAR VELOCITY FIELD}
\shortauthors{MOURI \& TANIGUCHI}

\received{2005 April 17}
\begin{document}

\title{HIERARCHICAL OBJECT FORMATION IN THE PECULIAR VELOCITY FIELD\thanks{This version corrects errors in equations (\ref{eq19}), (\ref{eq20}), (\ref{eq22}), and (\ref{eq25}) and Figures \ref{f2}$a$ and \ref{f3} of the published version (ApJ, 634, 20 [2005]). The errors are also corrected in the erratum (ApJ, 659, 1792 [2007]).}}

\author{HIDEAKI MOURI}
\affil{Meteorological Research Institute, Nagamine, Tsukuba 305-0052, Japan}

\and

\author{YOSHIAKI TANIGUCHI}
\affil{Astronomical Institute, Graduate School of Science, Tohoku University, Aoba, Sendai 980-8578, Japan; tani@astr.tohoku.ac.jp}

\notetoeditor{You may find the expression ``$10^2$'' etc. in our order-of-magnitude discussion. Please do not replace it with, e.g., ``100''.}

\notetoeditor{Figures 1-3 are too wide for a single column.}

\begin{abstract}

Using the initial peculiar-velocity field, we analytically study the hierarchical formation of gravitationally bound objects. The field is smoothed over a scale that corresponds to the mass of a given class of objects. Through the Zel'dovich approximation, the smoothed field determines how the objects cluster together to form a new class of more massive objects. The standard cosmological parameters lead to the evolution of primordial clouds with $\le 10^6\,M_{\sun}$ $\rightarrow$ galaxies with $10^{12}\,M_{\sun}$ $\rightarrow$ clusters of galaxies with $10^{15}\,M_{\sun}$ $\rightarrow$ superclusters of galaxies with $10^{16}\,M_{\sun}$. The epochs obtained for the formation of these classes of objects are consistent with observations.
\end{abstract}


\section{INTRODUCTION}
\label{s1}

For the formation of gravitationally bound objects such as galaxies and clusters of galaxies, the current paradigm is hierarchical formation based on the cold dark matter model. The cold dark matter induces significant small-scale fluctuations in the initial field. Since the first objects arise from these initial fluctuations, they are not so massive. They merge with one another via gravitational clustering in a hierarchy, evolving successively into more massive objects. We analytically discuss some statistical properties of this hierarchical object formation via the gravitational clustering.

The existing analytical studies are based on the density field, following the pioneering works of Press \& Schechter (1974; see also Peebles 1980), Bardeen et al. (1986), and Bond et al. (1991). They smoothed the initial density field over a scale that corresponds to a given class of objects. This is because, even when objects are formed at a scale, the larger-scale density field retains its initial pattern, according to the Eulerian linear theory. The smoothed initial density field was used to study the gravitational clustering of those objects.

Our study is based on the peculiar-velocity field, which is expected to offer new information on the gravitational clustering. Over scales where objects have not been formed, the peculiar-velocity field retains its initial pattern, according to the Lagrangian linear theory, i.e., the Zel'dovich approximation (ZA; Zel'dovich 1970). The initial fluctuations are random Gaussian. For such fluctuations, there are statistical measures (Rice 1954, \S3). We apply these measures to the smoothed initial peculiar-velocity field.

The gravitational clustering reflects the cosmological parameters. We use their latest standard values: the Hubble constant $H_0 = 70$ km s$^{-1}$ Mpc$^{-1}$ ($h = 0.7$), the cosmological constant $\Omega _{\Lambda} = 0.7$, the matter density $\Omega _{\rm m} = 0.3$, and the baryon density $\Omega _{\rm b} = 0.05$  (Spergel et al. 2003).

The definitions and equations are summarized in \S\ref{s2}. We discuss the general features of the hierarchical object formation via the gravitational clustering in \S\ref{s31}. The characteristic mass scales are discussed in \S\ref{s32}. The epochs for the formation of objects are discussed in \S\ref{s33}. We conclude with remarks in \S\ref{s4}.

\section{DEFINITIONS AND EQUATIONS}
\label{s2}

\subsection{Eulerian Description and Linear Theory}
\label{s21}

The basic assumption for the following equations is that the density field is isotropic and homogeneous (e.g., Peebles 1980, \S10, 27, and 41). We use comoving coordinates {\boldmath $x$}, i.e., physical coordinates divided by the scale factor. The scale factor $a(t)$ is normalized to $a(t_0) = 1$ at the present time $t = t_0$.

The mass density $\rho (\mbox{\boldmath $x$},t)$ and its spatial average $\langle \rho \rangle = 3 \Omega _{\rm m} H_0^2 / 8 \pi G$ define the density contrast $\delta (\mbox{\boldmath $x$},t)$ as $\rho / \langle \rho \rangle -1$. Here $\langle \cdot \rangle$ denotes an average. The relation to the Fourier transform $\tilde{\delta}(\mbox{\boldmath $k$},t)$ is
\begin{equation}
\label{eq1}
\delta (\mbox{\boldmath $x$},t) 
=
\frac{1}{(2 \pi)^{3/2}}
\int \tilde{\delta} (\mbox{\boldmath $k$},t) 
\exp ( i \mbox{\boldmath $k$} \cdot \mbox{\boldmath $x$}) d \mbox{\boldmath $k$} .
\end{equation}
For the wavenumber $k = \vert \mbox{\boldmath $k$} \vert$, the power spectrum $P(k,t)$ is
\begin{equation}
\label{eq2}
\langle \tilde{\delta} (\mbox{\boldmath $k$},t) 
        \tilde{\delta} (\mbox{\boldmath $k$}',t)^{\ast} \rangle
=
(2 \pi )^3 \delta _{\rm D} (\mbox{\boldmath $k$} - \mbox{\boldmath $k$}') P(k,t) .
\end{equation}
Here $\tilde{\delta} (\mbox{\boldmath $k$},t)^{\ast}$ is the complex conjugate of $\tilde{\delta} (\mbox{\boldmath $k$},t)$, and $\delta _{\rm D}(\mbox{\boldmath $x$})$ is Dirac's delta function.

So far as in the linear regime where the linear theory is valid, the density contrast and power spectrum grow in a self-similar manner:
\begin{equation}
\label{eq3}
\delta (\mbox{\boldmath $x$},t) \propto D(t)
\
\mbox{and}
\
P(k,t) = D(t)^2 P_{\rm lin}(k) \propto D(t)^2.
\end{equation}
The linear growth factor $D(t)$ is
\begin{equation}
\label{eq4}
D(a) = \frac{5 \Omega _{\rm m}}{2}
       \frac{H(a)}{H_0}
       \int ^a_0
       \frac{H_0^3da'}{a'^3 H(a')^3},
\end{equation}
with the Hubble parameter
\begin{equation}
\label{eq5}
H(a) = H_0 \left(  \frac{\Omega _{\rm m}}{a^3}
                 + \Omega _{\rm \Lambda}
                 + \frac{1- \Omega _{\rm m} - \Omega _{\rm \Lambda}}{a^2}
           \right) ^{1/2}.
\end{equation} 
Together with equations  (\ref{eq1}) and (\ref{eq3}), the continuity equation $\partial _t \rho + \mbox{div}(\rho \mbox{\boldmath{$v$}}) = 0$ yields the peculiar velocity
\begin{equation}
\label{eq6}
\mbox{\boldmath $v$}(\mbox{\boldmath $x$},t)
=
\frac{1}{(2 \pi)^{3/2}}
\frac{\dot{D}(t)}{D(t)}
\int  \frac{ i \mbox{\boldmath $k$}}{k^2}
      \tilde{\delta}(\mbox{\boldmath $k$},t)
      \exp ( i \mbox{\boldmath $k$} \cdot \mbox{\boldmath $x$}) 
      d \mbox{\boldmath $k$} .
\end{equation}
Here $\dot{D}(t)$ is the time derivative of the linear growth factor $D(t)$. The factor $\mbox{\boldmath $k$} /k^2$ implies that the peculiar velocity is dominated by larger scales than the density contrast. This property of the peculiar velocity is essential to our discussion in \S\ref{s3}.

The two-point correlation of the density contrast $\langle \delta (\mbox{\boldmath $x$},t) \delta (\mbox{\boldmath $x$} + \mbox{\boldmath $r$},t) \rangle$ at the scale $r = \vert \mbox{\boldmath $r$} \vert$ is computed from the power spectrum:
\begin{equation}
\label{eq7}
\langle \delta (\mbox{\boldmath $x$},t) \delta (\mbox{\boldmath $x$} + \mbox{\boldmath $r$},t) \rangle
=
4 \pi D(t)^2 \int ^{\infty}_{0} k^2 \, j_0 (kr) \, P_{\rm lin}(k) dk .
\end{equation}
For the pairwise peculiar velocity, i.e., the relative velocity that is parallel to the separation vector {\boldmath $r$}, the two-point correlation $\langle u (\mbox{\boldmath $x$},t) u (\mbox{\boldmath $x$} + \mbox{\boldmath $r$},t) \rangle$ is
\begin{equation}
\label{eq8}
\langle u (\mbox{\boldmath $x$},t) u (\mbox{\boldmath $x$} + \mbox{\boldmath $r$},t) \rangle
=
4 \pi\dot{D}(t) ^2 
\int ^{\infty}_{0}
\left[ j_0 (kr) - \frac{2 j_1 (kr)}{kr} \right] P_{\rm lin}(k) dk 
\end{equation}
(Monin \& Yaglom 1975, \S12; G\'orski 1988). Here $j_0(x)$ and $j_1(x)$ are first-kind spherical Bessel functions
\begin{equation}
\label{eq9}
j_0 (x) = \frac{\sin (x)}{x} 
\
\mbox{and}
\
j_1(x) = \frac{\sin (x)}{x^2} - \frac{\cos (x)}{x}.
\end{equation}
We defined the two-point correlations in terms of the linear growth factor $D(t)$ because we use them only in the linear regime.

\subsection{Lagrangian Description and ZA}
\label{s22}

Consider a mass element that occupies the position {\boldmath $x$}$_{\rm in}$ at the initial time $t_{\rm in}$. This initial or Lagrangian position {\boldmath $x$}$_{\rm in}$ is related to the position $\mbox{\boldmath $x$}( \mbox{\boldmath $x$}_{\rm in},t)$ at the time $t$ as
\begin{equation}
\label{eq10}
\mbox{\boldmath $x$}(\mbox{\boldmath $x$}_{\rm in},t) 
=
 \mbox{\boldmath $x$}_{\rm in}
+\frac{D(t)}{\dot{D}(t_{\rm in})} 
 \mbox{\boldmath $v$}(\mbox{\boldmath $x$}_{\rm in},t_{\rm in}) ,
\end{equation}
according to the ZA (Zel'dovich 1970; see also Shandarin \& Zel'dovich 1989). The initial velocity $\mbox{\boldmath $v$}(\mbox{\boldmath $x$}_{\rm in},t_{\rm in})$ is determined on the Eulerian linear theory (eq. [\ref{eq6}]).

Even when the Eulerian linear theory becomes invalid, the ZA is valid (e.g., Yoshisato, Matsubara, \& Morikawa 1998). Moreover, as in equation (\ref{eq10}), the ZA describes the motion of mass elements in gravitational clustering.

The ZA becomes invalid when the trajectories of mass elements starting from different positions intersect one another. The ZA mass elements pass away without any interaction, which occurs in practice and leads to formation of a gravitationally bound object (Gurbatov, Saichev, \& Shandarin 1989; Shandarin \& Zel'dovich 1989). We thereby define the object formation as this trajectory intersection. Since the trajectory intersection proceeds from small to large scales (e.g., Yoshisato, Morikawa, \& Mouri 2003), the object formation proceeds from low to high masses in accordance with the current paradigm (\S\ref{s1}). The usual definition for the object formation is that the density contrast $\delta (\mbox{\boldmath $x$},t)$ reaches approximately unity in the Eulerian linear theory (e.g., Press \& Schechter 1974; Bardeen et al. 1986; Bond et al. 1991). This is consistent with our definition because the true density contrast is very large in both the cases.

\subsection{Smoothing}
\label{s23}

The smoothing is based on a Gaussian window function, which is localized in the real and wavenumber spaces and is convenient. If the smoothing scale is $R$, we equivalently use the new power spectrum
\begin{equation}
\label{eq11}
P_{\rm lin}(R|k) = \exp (- k^2R^2) P_{\rm lin}(k) .
\end{equation}
The relation between the smoothing scale $R$ and the mass $M_R$ of the corresponding class of objects depends on the objects themselves (Bardeen et al. 1986; Bond et al. 1991). Since we are to assume that the objects have collapsed, we approximate as
\begin{equation}
\label{eq12}
M_R = \frac{4 \pi}{3} R^3 \langle \rho \rangle .
\end{equation}
This approximation is in accordance with the definition of characteristic mass scales in \S\ref{s32} (eq. [\ref{eq24}]). The mass $M_R$ includes both baryonic and dark matters.

When we are interested in a class of objects with a mass $M_R$, the initial peculiar-velocity field $\mbox{\boldmath $v$}(\mbox{\boldmath $x$},t_{\rm in})$ is smoothed over the scale $R$. The smoothed field determines the motion of those objects via the ZA (Mann, Heavens, \& Peacock 1993). Although the ZA is invalid over scales where objects have been formed, it is valid over the larger scales. They are in the linear or so-called quasi-linear regime. The ZA on a smoothed field, i.e., the truncated ZA, is actually known as a better approximation than the ZA on the original field (e.g., Coles, Melott, \& Shandarin 1993; Melott, Pellman, \& Shandarin 1994).

The smoothing over a scale $R$ and the subsequent application of the ZA imply that we consider not the initial time but a time when objects with the mass $M_R$ have been formed. A larger smoothing scale corresponds to a later time because the object formation proceeds from low to high masses (\S\ref{s1}).

The smoothing is also necessary to obtain statistics such as the root-mean-square density contrast (\S\ref{s31}). Without the smoothing, in the cold dark matter model, there are significant small-scale fluctuations that cause divergence of those statistics (Peebles 1980, \S26 and 96; Bardeen et al. 1986).

\subsection{Initial Power Spectrum}
\label{s24}

The density contrast and peculiar velocity at the initial time are not only isotropic and homogeneous but also random Gaussian. Their statistics are uniquely determined by the initial power spectrum
\begin{equation}
\label{eq13}
P(k,t_{\rm in}) = D(t_{\rm in})^2 P_{\rm lin}(k)
\
\mbox{with}
\
P_{\rm lin}(k) = T(k)^2 Ck.
\end{equation}
The transfer function $T(k)$ is from the cold dark matter model of Bardeen et al. (1986):
\begin{equation}
\label{eq14}
T(k) = \frac{\ln (1+2.34q)}{2.34q}
       \left[ 1+3.89q+(16.1q)^2+(5.46q)^3+(6.71q)^4 \right] ^{-1/4} .
\end{equation}
Here the normalized wavenumber $q$ is defined with the shape parameter $\Gamma$ in Sugiyama (1995):
\begin{equation}
\label{eq15}
q = \frac{k}{\Gamma h} \mbox{ Mpc}^{-1}
\
\mbox{with}
\
\Gamma = \Omega _{\rm m} h 
         \exp \left[- \Omega _{\rm b} \left( 1+\frac{\sqrt{2h}}{\Omega _{\rm m}} \right) \right].
\end{equation}
We use the scale-invariant Harrison-Zel'dovich spectrum $Ck$. The normalization constant $C$ is set to be $1 \times 10^5$ on the basis of the root-mean-square mass contrast within the radius $8\,h^{-1}$\,Mpc observed for the present time, 0.9 (Spergel et al. 2003; see also Peebles 1980, \S26).

\subsection{Statistics of Random Gaussian Field}
\label{s25}

For a homogeneous random Gaussian signal $y(x)$ in a one-dimensional space, there are useful statistical measures (Rice 1954, \S3). The mean interval between successive zeros is
\begin{equation}
\label{eq16}
\lambda
=
\pi \left[ \frac{ \langle y^2 \rangle }{ \langle (dy/dx)^2 \rangle } \right] ^{1/2} 
=
\pi \left[ - \frac{\xi (0)}{ d^2 \xi (0) /dr^2} \right] ^{1/2} .
\end{equation}
Here $\xi (r)$ is the two-point correlation $\langle y(x) y(x+r) \rangle$. The scale $\lambda$ is equivalent to the Taylor microscale that is often used to characterize turbulence (e.g., Monin \& Yaglom 1975, \S12 and 15). Likewise, the mean interval between successive extrema is
\begin{equation}
\label{eq17}
\eta
=
\pi \left[ \frac{ \langle (dy/dx)^2      \rangle }
                { \langle (d^2 y/dx^2)^2 \rangle } \right] ^{1/2}
=
\pi \left[ - \frac{ d^2 \xi (0) /dr^2}{ d^4 \xi (0) /dr^4} \right] ^{1/2} .
\end{equation}
The probability that $y(x)$ passes through a value $y_{\ast}$ with a positive slope in the range $(x, x+\delta x)$ is
\begin{equation}
\label{eq18}
p(y_{\ast}) \delta x
=
\frac{1}{2\lambda} \exp \left[ -\frac{y_{\ast}^2}{2 \langle y ^2 \rangle} \right] \delta x
=
\frac{1}{2\lambda} \exp \left[ -\frac{y_{\ast}^2}{2 \xi (0)}              \right] \delta x
.
\end{equation}
Since $\lambda$ is the mean size of regions where $y(x)$ has the same sign and there are plus and minus signs, the factor $\exp (-y_{\ast}^2 /2 \langle y^2 \rangle )$ is the probability that $| y(x) |$ passes thorough a value $| y_{\ast} |$ with a positive slope in such a region. This is just the probability that the maximum of $|y(x)|$ exceeds the value $|y_{\ast}|$. We differentiate $\exp (- y_{\ast}^2 /2 \langle y^2 \rangle )$ to obtain $(|y_{\ast}| / \langle y^2 \rangle ) \exp (-y_{\ast}^2 / 2 \langle y ^2 \rangle )$ as the probability that the maximum takes the value $|y_{\ast}|$.

\section{DISCUSSION}
\label{s3}

Our discussion is limited to one-dimensional cuts of the three-dimensional peculiar-velocity and density fields. Only for such a one-dimensional signal, useful statistical measures exist (\S\ref{s25} and \ref{s32}). They have been nevertheless used in three-dimensional turbulence for decades (e.g., Monin \& Yaglom 1975, \S12 and 15) and are hence expected to yield crude but still useful information on three-dimensional process of the object formation.

\subsection{General Features}
\label{s31}

Figure \ref{f1} shows the peculiar velocity $v_x(R|x,t_{\rm in})$ and the density contrast $\delta (R|x,t_{\rm in})$ at the initial time $t_{\rm in}$ on a one-dimensional cut $x$ smoothed over the scale $R = 0.01$\,Mpc. They are normalized with their root-mean-square values. The peculiar velocity $v_x$ is in the direction of the one-dimensional cut.

The peculiar velocity and density contrast were obtained from the three-dimensional Fourier transforms. Their real and imaginary parts are mutually independent and random Gaussian. We assigned the Fourier transforms according to the initial power spectrum (eq. [\ref{eq2}]) and converted them to the real-space counterparts (eqs. [\ref{eq1}] and [\ref{eq6}]).

The density field has many peaks. They individually collapse into gravitationally bound objects. The peculiar-velocity field has larger-scale variations (\S\ref{s21}). Then, as predicted by the ZA and indicated by arrows, the density peaks cluster together. They form a new class of more massive objects, provided that the clustering velocity is high enough.\footnote{
The collapse of the individual density peaks into gravitationally bound objects occurs earlier than the clustering into a new class of more massive objects.}  Zeros of the peculiar velocity, i.e., stationary points of the clustering motion, are centers or edges of the clustering. There the density contrast is positively or negatively enhanced in a significant manner. We also see small-scale velocity variations, which correspond to collapse and merger of the individual density peaks during the clustering.

While the mean size of density peaks is equal to the mean interval $\eta _{\delta}$ between successive extrema of the density contrast, the mean size of overdense regions is equal to the mean interval $\lambda _{\delta}$ between successive zeros of the density contrast. The mean clustering scale is equal to the mean interval between successive ridges and troughs of the peculiar velocity, where the clustering motion is largest.\footnote{
This interval does not represent edge-to-edge sizes of the clustering regions. We thus ignore their outer envelopes.} 
This interval is equal to the mean interval $\lambda _u$ between successive zeros of the peculiar velocity. Together with the typical clustering velocity, the mean clustering scale $\lambda _u$ determines the typical clustering timescale.

\subsection{Characteristic Mass Scales}
\label{s32}

Using equation (\ref{eq8}) and its derivatives at $r = 0$, the length scales $\lambda _u(R)$ and $\eta _u(R)$ in equations (\ref{eq16}) and (\ref{eq17}) are obtained for the pairwise peculiar velocity at the initial time $t_{\rm in}$ smoothed over the scale $R$:
\begin{equation}
\label{eq19}
\lambda _{u}(R) = \pi \left[ \frac{5 \int ^{\infty}_0     P_{\rm lin}(R|k) dk}
                                  {3 \int ^{\infty}_0 k^2 P_{\rm lin}(R|k) dk} \right] ^{1/2} ,
\end{equation}
and
\begin{equation}
\label{eq20}
\eta _{u}(R) = \pi \left[ \frac{7 \int ^{\infty}_0 k^2 P_{\rm lin}(R|k) dk}
                               {5 \int ^{\infty}_0 k^4 P_{\rm lin}(R|k) dk} \right] ^{1/2} ,
\end{equation}
Likewise, using equation (\ref{eq7}) and its derivatives at $r = 0$, the length scales $\lambda _{\delta}(R)$ and $\eta _{\delta}(R)$ are obtained for the initial density contrast:
\begin{equation}
\label{eq21}
\lambda _{\delta}(R) = \pi \left[ \frac{3 \int ^{\infty}_0 k^2 P_{\rm lin}(R|k) dk}
                                       {  \int ^{\infty}_0 k^4 P_{\rm lin}(R|k) dk} \right] ^{1/2} ,
\end{equation}
and
\begin{equation}
\label{eq22}
\eta _{\delta}(R) = \pi \left[ \frac{5 \int ^{\infty}_0 k^4 P_{\rm lin}(R|k) dk}
                                    {3 \int ^{\infty}_0 k^6 P_{\rm lin}(R|k) dk} \right] ^{1/2} ,
\end{equation}
The scales $\eta _u(R)$ and $\lambda _{\delta}(R)$ are identical apart from a numerical factor. For the initial density contrast, we also obtain the correlation length $\kappa _{\delta}(R)$, i.e., the length scale characteristic of the large-scale correlation (e.g., Monin \& Yaglom 1975, \S12)
\begin{equation}
\label{eq23}
\kappa _{\delta}(R)
=
\frac{ \int ^{\infty}_0 \langle 
                        \delta (R|\mbox{\boldmath $x$},t_{\rm in}) 
                        \delta (R|\mbox{\boldmath $x$} + \mbox{\boldmath $r$},t_{\rm in}) \rangle dr}
     {                  \langle \delta (R|\mbox{\boldmath $x$},t_{\rm in}) ^2 \rangle }
=
\frac{ \pi \int ^{\infty}_0 k P_{\rm lin}(R|k) dk}{2  \int ^{\infty}_0 k^2 P_{\rm lin}(R|k) dk} .
\end{equation}
The correlation length is zero for the initial pairwise peculiar velocity because its two-point correlation is significantly negative at large scales (see below). Although $\lambda _u(R)$, $\eta _u(R)$, $\lambda _{\delta}(R)$, $\eta _{\delta}(R)$, and $\kappa _{\delta}(R)$ are defined at the initial time, they are valid throughout the linear regime where the peculiar velocity and density contrast remain random Gaussian.

The mass scales $M_{\lambda}$, $M_{\eta}$, and $M_{\kappa}$ are defined as the masses of spheres with the diameters $\lambda$, $\eta$, and $\kappa$:
\begin{equation}
\label{eq24}
M_{\lambda} = \frac{4 \pi}{3} \left( \frac{\lambda}{2} \right) ^3 \langle \rho \rangle ,
\
M_{\eta} =    \frac{4 \pi}{3} \left( \frac{\eta}{2}    \right) ^3 \langle \rho \rangle ,
\
M_{\kappa} =  \frac{4 \pi}{3} \left( \frac{\kappa}{2}  \right) ^3 \langle \rho \rangle .
\end{equation}
The mass scales $M_{\lambda _u(R)}$ and $M_{\eta _u(R)}$ are Lagrangian quantities and are hence valid so far as the ZA is valid. We regard the mean clustering mass $M_{\lambda _u(R)}$ as the mass for a new class of objects that are formed by objects with the mass $M_R$.

Figure \ref{f2}$a$ shows the mass scales for the pairwise peculiar velocity $M_{\lambda _u(R)}$ and $M_{\eta _u(R)}$ and those for the density contrast $M_{\lambda _{\delta}(R)}$, $M_{\eta _{\delta}(R)}$, and $M_{\kappa _{\delta}(R)}$ as a function of the smoothing mass $M_R$. The corresponding length scales are also shown.

Since the mean size of density peaks $\eta _{\delta}(R)$ is comparable to the smoothing scale $R$, the density peaks are the objects that correspond to the smoothing (Bardeen et al. 1986).\footnote{
The scale $\eta _u(R)$, which is considered to represent the motion associated with the collapse of the individual density peaks, is also comparable to the smoothing scale $R$.} The mean size of overdense regions $\lambda _{\delta}(R)$ is a few times greater. Since the mean clustering scale $\lambda _u(R)$ is far greater than the mean size of density peaks $\eta _{\delta}(R)$, the density peaks cluster together to form a new class of more massive objects. Since the mean clustering scale $\lambda _u(R)$ is also far greater than the density correlation length $\kappa _{\delta}(R)$, the clustering is independent of the individual density peaks.

The clustering is not significant for the length scales $\lambda _u(R) \ga 10^2$\,Mpc, where we see $\lambda _u(R) \simeq \lambda _{\delta}(R)$. While the individual overdense regions do not cluster together but collapse to single objects, the individual underdense regions expand to single voids. They form the so-called cellular structure where the objects constitute the cell walls. The two-point correlation of the pairwise peculiar velocity $\langle u (\mbox{\boldmath $x$},t_{\rm in}) u (\mbox{\boldmath $x$} + \mbox{\boldmath $r$},t_{\rm in}) \rangle$ at $r \ga  10^2$\,Mpc is correspondingly negative as in Figure \ref{f2}$b$.

Figure \ref{f2}$a$ is used to discuss the hierarchical formation of gravitationally bound objects, via successive smoothing of the peculiar-velocity field. The first objects are clouds with $10^6\,M_{\sun}$, which are considered to harbor the first stars in the universe (e.g., Bromm \& Larson 2004). The smoothing for $M_R = 10^6\,M_{\sun}$ yields the mean clustering mass $M_{\lambda _u(R)} \simeq 10^{12}\,M_{\sun}$, i.e., the typical mass for galaxies (e.g., Brainerd, Blandford, \& Smail 1996).\footnote{
The mean clustering mass $M_{\lambda _u(R)}$ is insensitive to the smoothing mass $M_R$ if it is less than about $10^6\,M_{\sun}$. This is because, as seen in equation (\ref{eq19}), the mean clustering scale $\lambda _u(R)$ is not so dependent on small-scale fluctuations.} 
Thus primordial clouds form galaxies. The smoothing for $M_R = 10^{12}\,M_{\sun}$ yields the mean clustering mass $M_{\lambda _u(R)} \simeq 10^{15}\,M_{\sun}$, i.e., the typical mass for clusters of galaxies (e.g., Merritt 1987). Thus galaxies form clusters of galaxies. The smoothing for $M_R = 10^{15}\,M_{\sun}$ yields the mean clustering mass $M_{\lambda _u(R)} \simeq 10^{16}\,M_{\sun}$. Thus clusters of galaxies form superclusters of galaxies. They are practically the most massive gravitationally bound objects because the clustering timescale for $M_{\lambda _u(R)} \gg 10^{16}\,M_{\sun}$ is too large (\S\ref{s33}). The mass scale $M_{\lambda _u(R)} \simeq 10^{16}\,M_{\sun}$ corresponds to the length scale $\lambda _u(R) \simeq 10^2$\,Mpc.

The smoothing mass $M_R$ in our above discussion was set to be the typical mass for an actual class of objects. Thus our discussion was analogous to observations, where stars are used to identify galaxies and galaxies are used to identify clusters of galaxies. It should be also noted that we considered a time when the class of objects with the mass $M_R$ have been formed (\S\ref{s23}).

Our success in reproducing distinct classes of gravitationally bound objects, i.e., galaxies, clusters of galaxies, and superclusters of galaxies, was due to the existence of the mean clustering scale $\lambda _u(R)$ or the mean clustering mass $M_{\lambda _u(R)}$ that is distinctly greater than the smoothing scale $R$ or the smoothing mass $M_R$, respectively. We underline that $\lambda _u(R)$ and $M_{\lambda _u(R)}$ are from the peculiar-velocity field. The density field does not offer information as to how the objects cluster together to form a new distinct class of more massive objects, at least at the level of the standard Eulerian linear theory. This is because, although the density field is related to the peculiar-velocity field, it is dominated by smaller-scale fluctuations (Fig. \ref{f1}; see also \S\ref{s21}).

\subsection{Formation Rate of New Objects}
\label{s33}

Within our framework of hierarchical object formation (\S\ref{s32}), we estimate the time for objects with a mass $M_R$ to form a new class of more massive objects. The mean clustering scale is $\lambda _u(R)$. The mean clustering mass, i.e., the mass for the new class of objects, is $M_{\lambda _u(R)}$.

The initial clustering velocity of objects with a mass $M_R$ is of the order of $\langle u(R|x,t_{\rm in})^2 \rangle ^{1/2}$. According to the ZA (eq. [\ref{eq10}]), the displacement of the objects by the time $t$ is $D(t)/\dot{D}(t_{\rm in})$ times $\nu \langle u(R|x,t_{\rm in})^2 \rangle ^{1/2}$. Here $\nu$ is a nondimensional parameter. When the trajectories of the objects intersect one another, the objects cluster together to form new objects (\S\ref{s22}). Since the mean clustering scale is $\lambda _u(R)$, the typical radius of the clustering region is $\lambda _u(R)/2$, which is regarded as the typical displacement. Using equations (\ref{eq8}), (\ref{eq16}), and (\ref{eq19}), we have 
\begin{equation}
\label{eq25}
D(t) = \frac{1}{4 \nu}
       \left[ \frac{5 \pi }{ \int ^{\infty}_0 k^2 P_{\rm lin} (R|k) dk} \right] ^{1/2} .
\end{equation}
For a value of the parameter $\nu$, equation (\ref{eq25}) determines the formation time of the new objects with the mass $M_{\lambda _u(R)}$.

Figure \ref{f3}$a$ shows the mass $M_{\lambda _u(R)}$ of new objects formed at redshift $z(t)+1 = a(t)^{-1}$ for the parameters $\nu = 1$--5. The mass is high at a low redshift. At each redshift, the mass is high for a large $\nu$ value, i.e., a large clustering velocity, which tends rare. Objects with $M_{\lambda _u(R)} \gg 10^{16}\,M_{\sun}$ are not formed because the object formation is suppressed at $z \la 0$. This suppression is due to the accelerated expansion of the universe. The increase of the linear growth factor $D(t)$ is no longer significant at $z \la 0$ as indicated by the dotted curve in Figure \ref{f3}$b$.

Given the mass $M_R$, equation (\ref{eq25}) determines the parameter $\nu$ as a function of the time $t$ for the formation of new objects with the mass $M_{\lambda _u(R)}$. The parameter $\nu$ represents the maximum velocity in the clustering region. Then its probability $p(\nu) \delta \nu$ is $\nu \exp (- \nu ^2 /2) \delta \nu$ (\S\ref{s25}). Using the relation $\delta \nu / \nu = \delta D /D$, which is from equation (\ref{eq25}), the probability is rewritten as
\begin{equation}
\label{eq26}
p( \nu ) \delta \nu
=
\nu ^2 \exp \left(- \frac{\nu ^2}{2} \right) \frac{\delta D}{D} .
\end{equation}
Equations (\ref{eq25}) and (\ref{eq26}) determine the formation rate of the new objects.

Figure \ref{f3}$b$ shows the formation rates as a function of redshift $z(t)+1$. For galaxies with $M_{\lambda _u(R)} \simeq 10^{12}\,M_{\sun}$, their formation by primordial clouds with $M_R = 10^6\,M_{\sun}$ starts at $z \simeq 10$ and is maximal at $z \simeq 3$. For clusters of galaxies with $M_{\lambda _u(R)} \simeq 10^{15}\,M_{\sun}$, their formation by galaxies with $M_R = 10^{12}\,M_{\sun}$ starts at $z \simeq 2$ and is maximal at $z \simeq 0$. This maximum corresponds to the saturation of the linear growth factor $D(t)$. For superclusters of galaxies with $M_{\lambda _u(R)} \simeq 10^{16}\,M_{\sun}$, their formation by clusters with $M_R = 10^{15}\,M_{\sun}$ starts at $z \la 0$.

The formation rates in Figure \ref{f3}$b$ are consistent with observations. Since the star formation rate integrated over galaxies has a broad peak at $z \simeq 4$--1 and declines to its present value (e.g., Heavens et al. 2004; Taniguchi et al. 2005), the typical galaxy mass is assembled by $z \simeq 1$. The number density of clusters of galaxies with $\ga 10^{15}\,M_{\sun}$ increases by a factor of 10 from $z \simeq 1$ to $z \simeq 0$ (Bahcall \& Bode 2003). Since the present galaxy distribution is almost uniform over scales above 40\,Mpc (e.g., Kurokawa, Morikawa, \& Mouri 2001; Jones et al. 2004), superclusters of galaxies have not undergone significant gravitational clustering at $z \simeq 0$.

There are analytical studies of the formation rate based on the density field (e.g., Lacey \& Cole 1993; Kitayama \& Suto 1996). The formation rate was obtained by integrating the transition rate from a mass $M$ to another $M'$. Since the transition rate diverges in the limit $M \rightarrow M'$, the formation of an object was defined as the assembly of a half of its mass. This definition is artificial and also does not correspond to the situation that objects cluster together to form a new distinct class of more massive objects. The density field does not offer the required information (\S\ref{s32}). We prefer our approach based on the peculiar-velocity field.

\section{CONCLUDING REMARKS}
\label{s4}

\begin{figure}
\begin{center}
\resizebox{12cm}{!}{\includegraphics*[3cm,7cm][18cm,20cm]{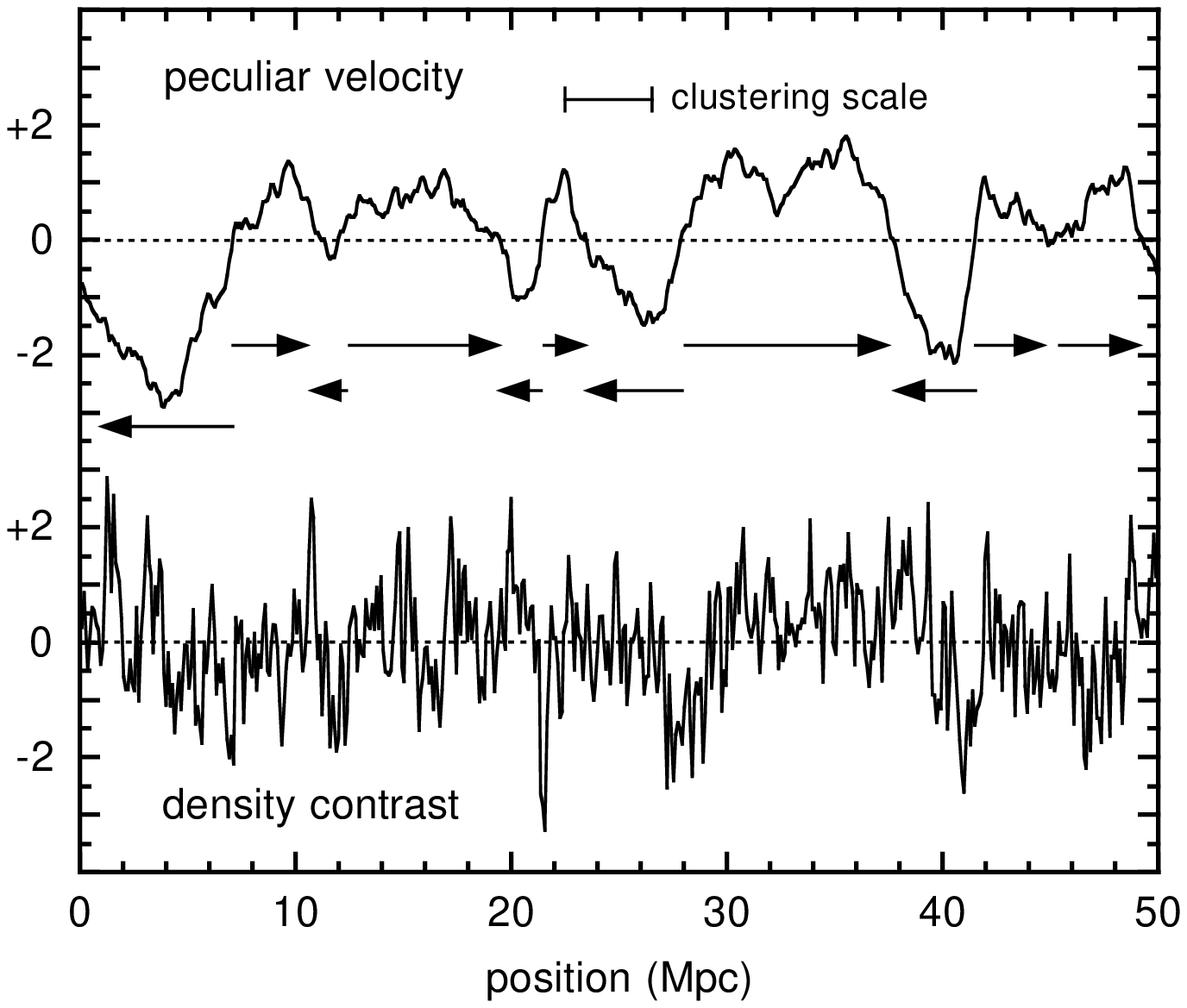}}

\caption[Fig1]{Peculiar velocity $v_x(R|x,t_{\rm in})$ and density contrast $\delta (R|x,t_{\rm in})$ at the initial time $t_{\rm in}$ on a one-dimensional cut $x$ smoothed over the scale $R = 0.01$\,Mpc. They are normalized with their root-mean-square values. The velocity $v_x$ is in the direction of the one-dimensional cut. The arrows indicate the clustering motion. We also indicate an example of the clustering scale. \label{f1}}

\end{center}
\end{figure}

\begin{figure}
\begin{center}

\resizebox{12cm}{!}{\includegraphics*[2cm,8cm][20cm,25cm]{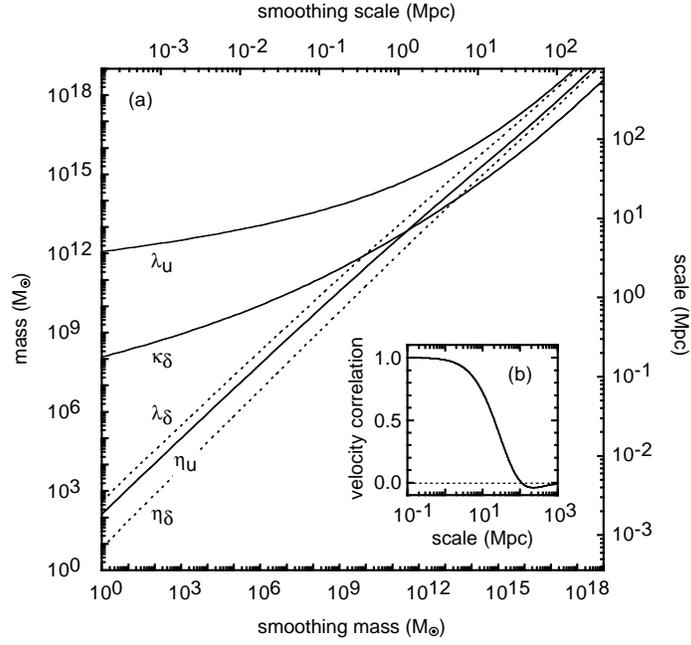}}

\caption[Fig2]{($a$) Mass scales for the pairwise peculiar velocity $M_{\lambda _u(R)}$ and $M_{\eta _u(R)}$ as well as mass scales for the density contrast $M_{\lambda _{\delta}(R)}$, $M_{\eta _{\delta}(R)}$, and $M_{\kappa _{\delta}(R)}$ as a function of the smoothing mass $M_R$. The corresponding length scales are also shown. ($b$) Two-point correlation of the pairwise peculiar velocity $\langle u (\mbox{\boldmath $x$},t_{\rm in}) u (\mbox{\boldmath $x$} + \mbox{\boldmath $r$},t_{\rm in}) \rangle$ normalized with $\langle u (\mbox{\boldmath $x$},t_{\rm in}) ^2 \rangle$. \label{f2}}

\end{center}
\end{figure}

\begin{figure}
\begin{center}

\resizebox{12cm}{!}{\includegraphics*[2cm,8cm][20cm,25cm]{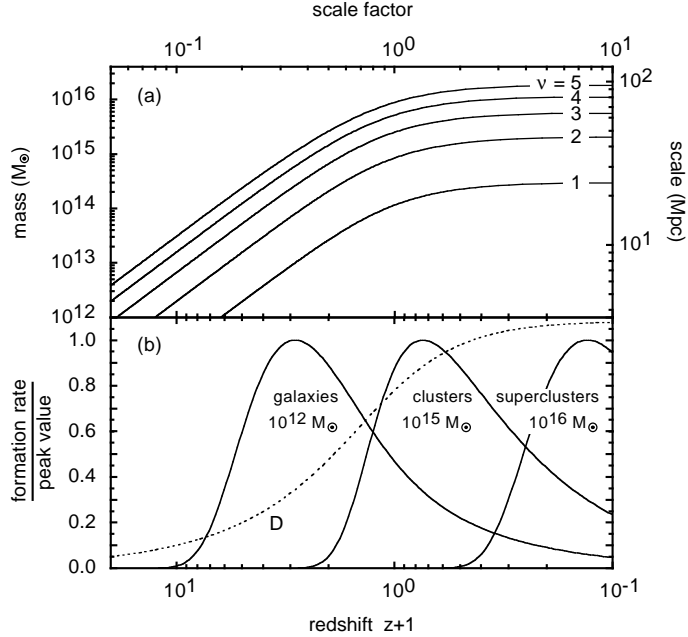}}
\caption[Fig3]{($a$) Mass $M_{\lambda _u(R)}$ of objects formed at redshift $z(t)+1$ for the clustering-velocity parameter $\nu = 1$--5. The length scale $\lambda _u(R)$ and the scale factor $a(t)$ are also shown. ($b$) Formation rates at redshift $z(t)+1$. We consider galaxies with $M_{\lambda _u(R)} \simeq 10^{12}\,M_{\sun}$ formed by primordial clouds with $M_R = 10^6\,M_{\sun}$, clusters of galaxies with $M_{\lambda _u(R)} \simeq 10^{15}\,M_{\sun}$ formed by galaxies with $M_R = 10^{12}\,M_{\sun}$, and superclusters of galaxies with $M_{\lambda _u(R)} \simeq 10^{16}\,M_{\sun}$ formed by clusters with $M_R = 10^{15}\,M_{\sun}$. The dotted curve denotes the linear growth factor $D(t)$. \label{f3}}

\end{center}
\end{figure}

The initial peculiar-velocity field was smoothed over a scale $R$ that corresponds to the mass $M_R$ of a class of gravitationally bound objects. Through the ZA, the smoothed field determines how the objects cluster together to form a new class of more massive objects (Fig. \ref{f1}). The mean clustering scale $\lambda _u(R)$ yields the mean clustering mass $M_{\lambda _u(R)}$, which is regarded as the mass for the new class of objects. We assumed the standard cosmological parameters and obtained the evolution of primordial clouds with $\le 10^6\,M_{\sun}$ $\rightarrow$ galaxies with $10^{12}\,M_{\sun}$ $\rightarrow$ clusters of galaxies with $10^{15}\,M_{\sun}$ $\rightarrow$ superclusters of galaxies with $10^{16}\,M_{\sun}$ (Fig. \ref{f2}$a$). Together with the clustering velocity, the mean clustering scale $\lambda _u(R)$ was also used to obtain the epochs for the formation of those classes of objects (Fig. \ref{f3}$b$). They are consistent with observations.

The peculiar-velocity field allowed us for the first time to analytically study the formation of distinct classes of gravitationally bound objects. This is because the peculiar-velocity field is dominated by large-scale fluctuations (Fig. \ref{f1}). There is the mean clustering scale $\lambda _u(R)$ that is distinctly greater than the smoothing scale $R$. Since it is difficult to obtain such a characteristic from the density field, the existing analytical studies based on the density field did not reproduce the distinct classes. There is even a model of continuous clustering, where galaxies and superclusters of galaxies are extremes of a continuous range of objects (Peebles 1980, \S4). This model does not appear so realistic. We also underline that the ZA used in our study is a better approximation than the Eulerian linear theory used in most of the existing studies.

Our study is based on one-dimensional cuts of the three-dimensional peculiar-velocity field. Although Bardeen et al. (1986) obtained useful statistical measures for a three-dimensional density field, there is no useful statistical measures for a three-dimensional velocity field. The advent of such measures is desirable. Then we could obtain more information on the peculiar-velocity field and hence the hierarchical formation of gravitationally bound objects.

\acknowledgments

The authors are grateful to M. Morikawa and A. Yoshisato for interesting discussion and to the referee for helpful comments.


\clearpage

\clearpage

\clearpage

\end{document}